\documentclass[aps,prb,twocolumn,floatfix,showpacs]{revtex4}

\usepackage{graphicx}
\usepackage{amsmath}
\usepackage[T1]{fontenc}
\usepackage[latin1]{inputenc}
\usepackage{times}
\usepackage[normalem]{ulem}

  \makeatletter
  \def\@dotsep{4.5}
  \makeatother

\newlength{\myVSpace}
\newcommand\xstrut{\raisebox{-.5\myVSpace}
  {\rule{0pt}{\myVSpace}}
}
\newcommand{\dif}{\mathrm{d}}
\newcommand{\Ang}{\mbox{\AA}}

\begin{document}

\title{A Periodic Genetic Algorithm with Real-Space Representation for 
Crystal Structure and Polymorph Prediction}
\author{N.L. Abraham}

\author{M.I.J. Probert}
\affiliation{Department of Physics, University of York, Heslington, York, 
YO10 5DD, United Kingdom}


\begin{abstract}
A novel Genetic Algorithm is described that is suitable for determining
the global minimum energy configurations of crystal
structures and which can also be used as a polymorph
search technique. This algorithm requires no prior assumptions about unit 
cell size, shape or symmetry, nor about the ionic configuration within the 
unit cell. This therefore enables true {\it ab initio} crystal structure 
and polymorph prediction. Our new algorithm uses a real-space
representation of the population members, and makes use of a novel
periodic cut for the crossover operation. Results on large
Lennard-Jones systems with FCC- and HCP-commensurate cells show
robust convergence to the bulk structure from a random initial
assignment and an ability to successfully discriminate between
competing low enthalpy configurations. Results from an {\it ab
initio} carbon polymorph search show the spontaneous emergence of both 
Lonsdaleite and graphite like structures.
\end{abstract}

\maketitle

\section{Introduction}

The prediction of crystal structures from first principles has long been 
recognized as one of the outstanding challenges in solid state physics 
\citep{Maddox88,VanDeWalle05}. The most recent methods of cluster expansion 
assume the lattice structure of the crystal \citep{BlumHWZ05,HartBWZ05}.
Good results for silicon have been shown using minima hopping 
\citep{Goedecker04}, but this method assumed the number of atoms and unit 
cell of the structures searched. In this communication we demonstrate a new 
method for unbiased {\it ab initio} crystal structure determination using a 
novel Genetic Algorithm which makes no assumptions of atom number, unit cell 
or lattice structure.

\begin{figure}[b]
\includegraphics*[scale=0.40]{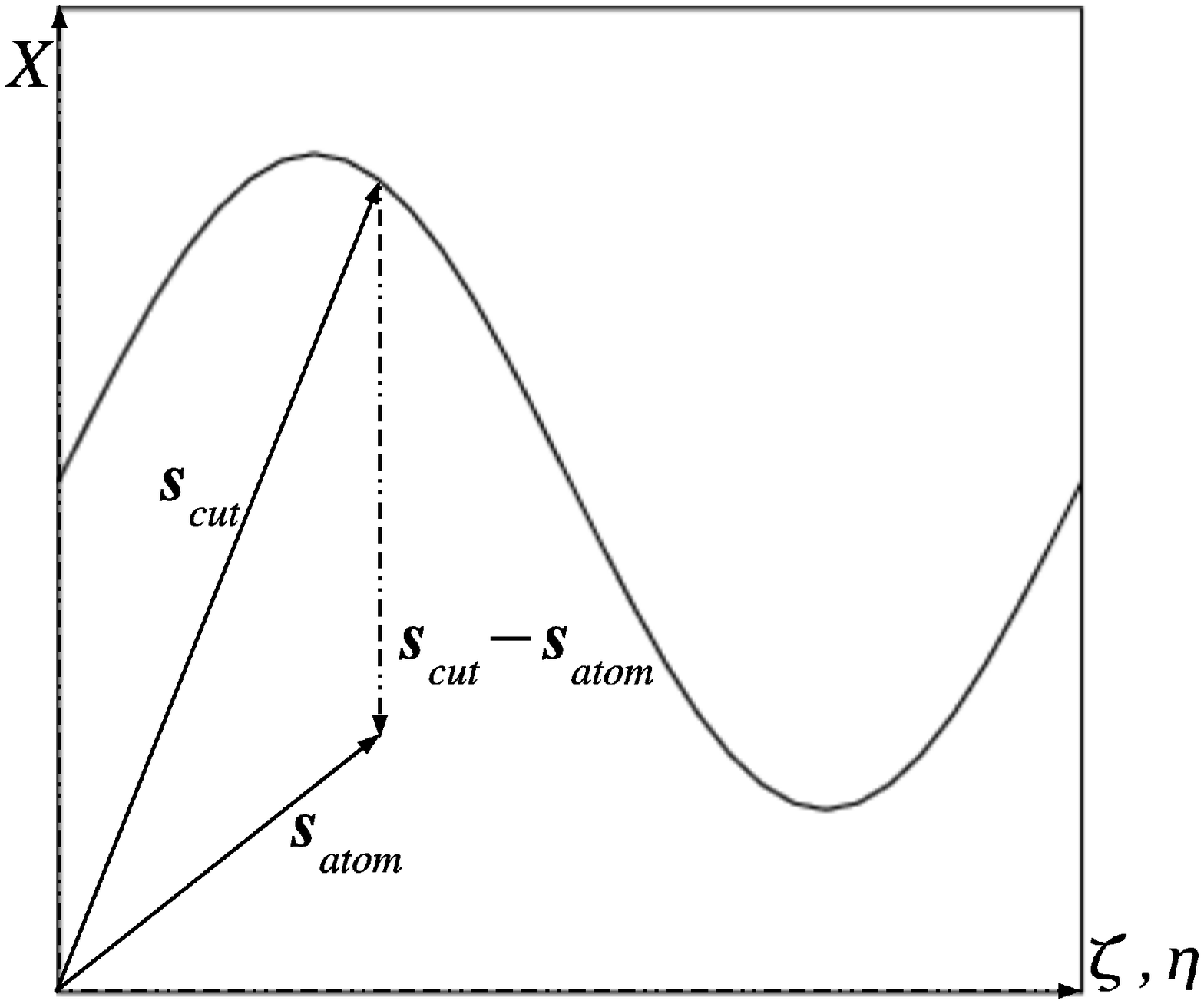}
\caption{Diagram showing crossover in a fractional representation. For each 
$\left(\zeta,\,\eta\right)=$ either $\left({\bf a},{\bf b}\right)$, 
$\left({\bf b},{\bf c}\right)$ or $\left({\bf c},{\bf a}\right)$ then 
${\bf X}=$ either ${\bf c}$, ${\bf a}$ or ${\bf b}$.}
\label{fig:cross_diag}
\end{figure}

Genetic Algorithms (GAs) were first developed by John Holland in
1975 \citep{Holland92}, and are stochastic global optimization
methods based on ``survival of the fittest''. This is a computational 
technique which is used to solve problems in which there are many potential
solutions, only a small number of which are optimal. We initialize our system 
with a number of random candidate solutions which are grouped together into a 
{\it population}, with each solution being a {\it member} of this population. 
There needs to be a way of determining the {\it fitness} of each member, i.e.
some way of telling which members are better or worse solutions to
the problem than the other members. It is this fitness function which defines 
the problem. Population members will be chosen or {\it selected}, based on 
their fitness, to become {\it parents} to produce {\it offspring} in a 
breeding procedure known as {\it crossover}. Crossover involves creating one 
or more offspring which are a combination of features from their parents. 
Each population member needs to be {\it encoded} or {\it represented} in some 
way such that crossover can be performed in a systematic way. For example, in 
the case of binary strings, two parents, ${11010011}$ and ${\bf 01011001}$ may 
be split in half and recombined to make two new offspring ${1101{\bf 1001}}$ 
and  ${{\bf 0101}0011}$. The offspring may be {\it mutated} after crossover, 
which involves making changes to offspring in a random way which could 
introduce beneficial aspects into the population. In the binary string case 
this most often involves changing a small percentage of the bits on the 
string. Using each member's fitness the population is {\it updated} by only 
allowing some population members to survive into the next {\it generation}, 
the rest of the population members will be discarded. The original formulation 
used a binary string representation, which is described in detail in 
\citet{Holland92}. In this study the problem is the determination of the 
optimal configuration of atoms, where the fitness function is a function of 
the enthalpy of the system. 

GAs were not widely used in the field of solid state physics due to
the representation issue, the binary nature of which was
insufficient to describe the complicated atomic and molecular
systems that are of interest. This changed in 1995 when \citet{DeavenH95} 
described a GA technique for clusters using a
representation which used the atomic coordinates of the systems
being studied. Crossover was performed by taking a planar cut
through the center of each parent and swapping halves to generate
offspring \citep{DeavenH95,Johnston03}. After crossover the
resulting structures of the resulting population members may be a
long way from equilibrium and so a quasi-Newton minimizer is used to
relax each structure to the minimum of the local basin of
attraction. This reduces the problem by simplifying the potential
energy surface that is searched by the GA. This method shows
excellent convergence for a number of complicated systems, and
recently studies of nanowires \citep{WangYWBZ01} and surfaces
\citep{ChuangCSWH04,ChuangCSPH05} have been performed using a
similar technique. The cluster calculations are performed using
non-periodic codes, but the nanowire and surface studies make
use of periodic boundary conditions (PBCs) in either one or two
dimensions, and a planar cut does not takes these PBCs into
account. Bulk binary-encoded GA studies have been presented by 
\citet{Woodley04} and real-space encoded studies of bulk 
molecular crystals have been reported by \citet{HarrisHCJ04} 
in which crossover was performed by randomly swapping of groups of parameters 
between population members, rather than considering the structural 
configuration of the atoms within the crystal. 

In this article, we demonstrate that for periodic GA studies using a 
periodic cut in the crossover operation is
superior to a planar one. The periodic cut is chosen to have the
same periodicity as the supercell of the population member, and
this reduces the discontinuities produced in the offspring
operation, which would cause extra work for the local minimizer
that is required in the GA scheme. Results show that a periodic
cut has a faster convergence than a planar one for large
systems. We also apply this technique to systems where the unit cell
of the solution is optimized as well as the atomic coordinates. Finally we 
show that GAs are suitable as a polymorph search technique.

\section{Method}

The crossover is performed in fractional coordinates, as described in figure 
\ref{fig:cross_diag}. The cut is defined by any periodic function with the 
same periodicity as the cell, 
${\bf f} \left({\bf s}^{\left(\zeta,\eta\right)}_{atom} \right)$,
where ${\bf s}^{\left(\zeta,\eta\right)}_{atom}$ is the fractional position 
vector for each atom along the 
$\left(\zeta,\eta\right)=\left({\bf a},{\bf b}\right)$, 
$\left({\bf b},{\bf c}\right)$ or $\left({\bf c},{\bf a}\right)$ directions.
This function gives a vector (in fractional coordinates) ${\bf s}_{cut}$ for 
each ${\bf s}_{atom}$ in the population member. The metric tensor 
$\uuline{g}=\uuline{h}^{\text{T}}\uuline{h}$, where 
$\uuline{h}=\left[{\bf a},{\bf b},{\bf c}\right]$ (in Cartesian coordinates), 
is used to calculate the product

\begin{equation}
{\alpha}_{cut}=\left({\bf s}_{cut}-{\bf s}_{atom}\right)^{\text{T}}\,
\uuline{g}\,{\bf X}
\label{eq:cross}
\end{equation}

where
${\bf X} =
\begin{cases}
\vspace{0.1cm}
{\bf a}&= \left[ \begin{smallmatrix}
1 \\ 0 \\ 0
\end{smallmatrix}  \right] \\
\vspace{0.1cm}
{\bf b}&= \left[ \begin{smallmatrix}
0 \\ 1 \\ 0
\end{smallmatrix}  \right] \\
{\bf c}&= \left[ \begin{smallmatrix}
0 \\ 0 \\ 1
\end{smallmatrix}  \right] \\
\end{cases}
\vspace{0.1cm}
$
in fractional coordinates (${\bf X} = {\bf c}$ when 
$\left(\zeta,\eta\right)=\left({\bf a},{\bf b}\right)$
etc.) and with the criterion

\begin{equation}
{\alpha}_{cut}
\begin{cases}
 > 0  \ & \ \text{the atom is \emph{``above''} (outside) the cut} \\
 \leq 0 \ & \ \text{the atom is \emph{``below''} (inside) the cut.} \\
\end{cases}
\label{eq:crit}
\end{equation}

As the cut is made in fractional coordinates, it does not
\emph{``know''} about the Cartesian shape of the cell, so this
technique allows two population members with different cells to be
bred during crossover, rather than being constrained to both
parents having the same supercell. The technique also allows the
cell size and shape to be evolved along with the crystal structure.

\begin{figure}[t]
\includegraphics*[scale=0.45]{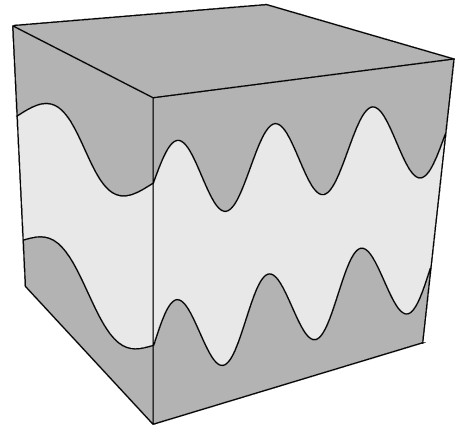}
\caption{Real-space representation of the periodic cuts in the crossover
operation. Different wavelengths and amplitudes can be used for
the cuts along the different cell directions. The cuts are calculated in 
fractional coordinates which allows crossover between parents with different 
cells. The dark gray sections represent one part of the cell, the light gray 
the other, and it is these parts that are swapped in crossover.} 
\label{fig:bulk_cross}
\end{figure}

Figure \ref{fig:bulk_cross} shows the crossover operation where two 
cuts are required in the cell. Every crossover operation
performed has an equal probability of being calculated with the
cuts made in reference to either the ${\bf a}$, ${\bf b}$ or ${\bf
c}$ directions. This ensures that none of the three co-ordinate directions 
is preferred over the other two, but also allows large areas of each
of the population members to be undisturbed by the cut. To ensure
that no one parent is preferred over the other, the center of
each of the cuts should be made one-quarter and three-quarters up
the chosen cell axis which gives approximately even mixing.

In the Deaven and Ho formulation, the plane of the cut was defined
by the creation of a random unit vector on the surface of a sphere
which was centered on the center of mass of the cluster. In our
method cuts made in different cell directions can have different
random wavelengths and amplitudes, although a maximum amplitude
should be defined so that the cut is contained within the
simulation supercell, and obviously the wavelength of the cut
cannot be longer than twice the cell vector in that direction, whilst 
any cut with a wavelength of less than half the
atomic separation will appear as a flat plane.

Fitness is determined by the relative enthalpy per atom of the
population members, and each population member was chosen for
crossover based on its fitness using roulette wheel selection
\citep{Johnston03}. Similar to \citet{ChuangCSWH04} the number of
atoms in each individual population member can be varied by
accepting all solutions after crossover, or if the number of atoms
needs to be constrained then solutions are rejected until
offspring are generated that have the correct number.

Roulette wheel selection can also be used in the update procedure,
which determines which members of the original population should
progress through to the next generation. In addition, the lowest energy
population member was guaranteed to be selected for update when
updating by this method. By allowing some higher energy members to
remain, this update procedure prevents the population from becoming 
stagnated. An alternative method for updating is to only allow the
lowest $M$ population members to proceed to the next generation,
from a super-population of $2M$ parents and off-spring. We will
refer to this method as the "simple" update scheme.

The way that mutation is performed is also important. This is
controlled by two quantities, $m_R$ and $m_A$. The mutation rate
is $m_R \in \left[0,1\right]$, and this determines the probability
that each atom will be mutated or not after crossover, before the 
local structure minimization procedure. Once an atom has been selected 
for mutation then it is randomly placed in a cubic box with sides of 
length $2m_A$ which has been centered on the atom's original position.

\section{Results}

\subsection{Results from the Empirical Lennard-Jones potential}

All calculations were performed by adding the above GA formulation
to the \emph{ab initio} planewave DFT code CASTEP \citep{Castep}, which 
has also been modified for ease of algorithm testing to allow the use of 
the empirical Lennard-Jones potential \cite{LennardJonesI25}

\begin{equation}
\label{eq:lj_pot}
V_{LJ}\left(r_{ij}\right) = 4\epsilon\left[
{\left(\frac{\sigma}{r_{ij}}\right)}^{12} - 
{\left(\frac{\sigma}{r_{ij}}\right)}^{6}\right]
\end{equation}
in the shifted-force formulation \citep{StoddardF73}
\begin{widetext}
\begin{equation}
\label{eq:lj_shift}
V^{SF}\left(r_{ij}\right) =
\begin{cases}
{V_{LJ}\left(r_{ij}\right) - V_{LJ}\left(r_{cut}\right) - 
\left(\frac{\dif V_{LJ}\left(r_{ij}\right)}{\dif r_{ij}}\right)_{\left(r_{ij}
=r_{cut}\right)}\left(r_{ij}-r_{cut}\right)}
\ \, & r_{ij} \leq r_{cut} \\
0 & r_{ij} > r_{cut} \\
\end{cases}
\end{equation}
\end{widetext}
which has a HCP ground state structure \citep{Pollack64} which is
almost degenerate with the FCC structure (energy difference from
HCP $+0.1\,\%$ \citep{KaneG40}. The energy difference between the
FCC and HCP supercells used in this study was $+0.072\,\%$, due to
the above formulation of the Lennard-Jones potential). The value
of $\sigma$ was set to $3.405\,\Ang$, $\epsilon$ was set to $120\,\mbox{K}$, 
and $r_{cut}$ was set to $2.5\sigma$. While the ground states are very close 
in energy, to switch from FCC to HCP four out of every six layers require a 
stacking fault.

\begin{figure}[b]
\includegraphics*[scale=0.80]{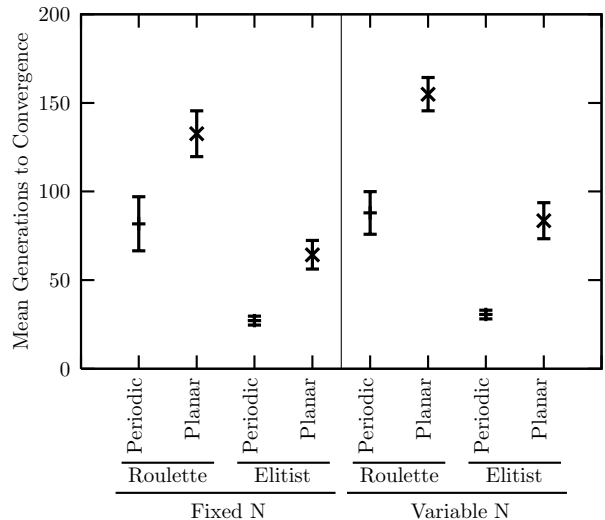}
\caption{Summary of the convergence times (in number of generations) for 
each variation of the method presented. For each run, either periodic or 
planar cuts were used, using either the roulette or the simple update 
scheme, and the number of atoms could either be kept fixed, or be allowed 
to vary.}
\label{fig:convergence}
\end{figure}

For the Lennard-Jones results we used a fixed supercell, but
allowed the number of atoms to either be fixed for each of the
population members, or be allowed to vary. We compared GA
minimization calculations using either the planar or periodic cuts
as described above. The number of population members was fixed at
$M=16$, and the initial number of atoms in each population member 
was set to $N=150$ using a hexagonal supercell, which
is commensurate with perfect FCC and HCP structures without stacking 
faults. The mutation rate, $m_R$, was set to $0.10$ ($10\,\%$) and the 
mutation amplitude, $m_A$, was set to $2.5\Ang$. The initial configuration 
of the population members is totally randomized, then minimized with the 
local minimizer before proceeding. A total of $200$
generations was run for each simulation.

In total 15 simulations were performed from a random start for each of the 
eight combinations of either fixed or variable number of atoms, using either 
the roulette wheel or simple update scheme, and with crossover performed 
using either periodic cuts or a planar cut. A summary of the convergence 
times is shown in figure \ref{fig:convergence}. 

\subsubsection{Empirical Lennard-Jones bulk studies with a fixed 
number of atoms.}
\label{sec:fixedN}

For these studies the number of atoms was kept fixed at 150 during the whole 
of the simulation. However, none of the 60 calculations resulted in 
minimization down to FCC or HCP stacking. The simple update scheme was much 
faster at reaching convergence, and using periodic cuts was much faster than 
using a planar cut when using either update scheme.

\subsubsection{Empirical Lennard-Jones bulk studies with a variable 
number of atoms.}

Results from \citet{ChuangCSWH04} showed that allowing the number
of atoms to vary helped convergence. Our results also show this. The use of
periodic cuts using the roulette wheel update scheme, or the planar cut
using the roulette wheel or simple update scheme, allowed the system to be
minimized to a defect-free ground state structure. Periodic cuts were faster 
to convergence than a planar cut, as shown above.

\begin{figure}[t]
\includegraphics*[scale=0.78]{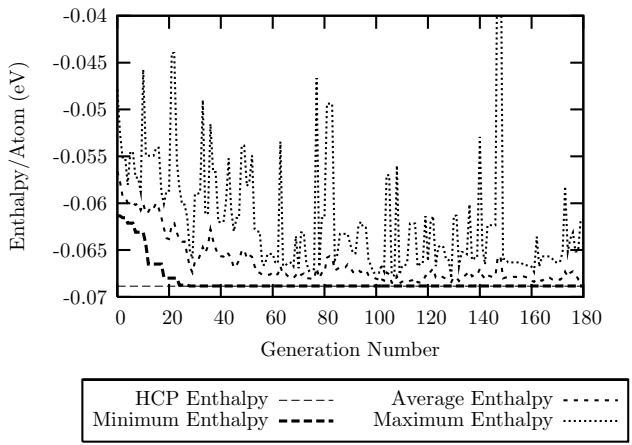}
\caption{Typical results from a 150 atom (variable), 16 population member
calculation starting from an initial random configuration with a mutation 
rate of $10\%$ and mutation amplitude of $2.5\Ang$, using roulette wheel 
selection in the update procedure. Periodic cuts were used and the system 
converged in 29 generations to a structure with 150 atoms and a local minimum 
enthalpy $+0.024\,\%$ above the HCP minimum. 
}
\label{fig:bulk_graph2a}
\end{figure}

\begin{figure}[b]
\includegraphics*[scale=0.25]{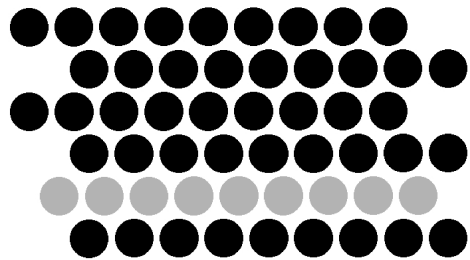}
\caption{Side on view of minimized structure from figure 
\ref{fig:bulk_graph2a}, looking down the $\left[0\bar{1}1\right]$ direction. 
The colors show the mixture of FCC (gray) and HCP (black) stacking.}
\label{fig:side_view_fnf}
\end{figure}

We found that the system did not converge into a perfect lattice structure 
without allowing for variable atom number. Figure \ref{fig:bulk_graph2a} shows 
a typical set of results, using periodic cuts with roulette wheel selection 
for update. In this case the system converged in 29 generations to the 
structure shown in figure \ref{fig:side_view_fnf}. This configuration is an 
FCC-HCP hybrid with an energy difference of $+0.024\,\%$ from the HCP ground 
state, which is due to a single FCC plane stacking fault. Similar structures 
were also found using a planar cut, but with longer convergence times.

\begin{figure}[t]
\includegraphics*[scale=0.50]{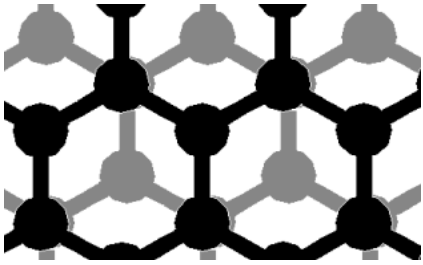}
\caption{Top view of structure 1 of the carbon structures (graphite-like) 
showing the layer stacking. The top layer is colored black and the layer 
below gray for clarity.}
\label{fig:c4_01}
\end{figure}

\begin{figure}[b]
\includegraphics*[scale=0.50]{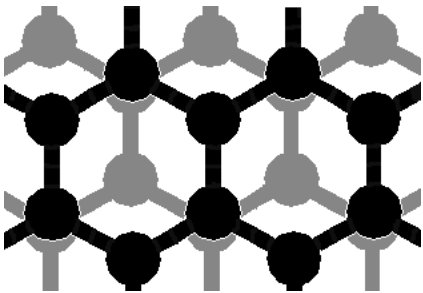}
\caption{Top view of structure 5 of the carbon structures (graphite-like) 
showing the mismatch in the layer stacking. The top layer is colored black 
and the layer below gray for clarity.}
\label{fig:c4_05}
\end{figure}

\subsection{{\it Ab initio} carbon polymorph studies with a variable 
supercell.}

An interesting alternative application of our GA is in the field of polymorph 
prediction. For example, there has been considerable interest recently in 
carbon polymorphs. Systematic {\it ab initio} searches been made of 
$sp^{3}$-hybridized structures with four atoms per unit cell 
\citep{StrongPMTW04} and of $sp^{2}$-hybridized structures with four or six 
atoms per unit cell \citep{WinklerPMT01,WinklerPMKT99}. 

\begin{figure}[t]
\includegraphics*[scale=0.50]{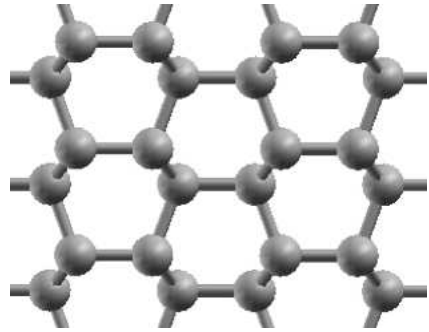}
\caption{View of structure 7 of the carbon structures (Lonsdaleite) 
looking down the $\left[110\right]$ direction.}
\label{fig:c4_07}
\end{figure}

Such calculations either require a graph-theoretical enumeration of possible 
configurations, resulting in an exponentially growing search space as the 
number of atoms per unit cell increases, or serendipity. By contrast, our GA 
approach is not restricted to any particular form of bonding or chemical 
intuition, and is very efficient at searching high dimensional spaces. As an 
application of our GA to polymorph determination we therefore chose
to study the four atom per unit cell carbon polymorphs.

All calculations were performed using density functional theory (DFT) to 
treat the electrons for a given configuration of ions so no assumptions were 
made about the nature of the underlying bonding. DFT has been shown to be 
very accurate for the calculation of atomic configurations many times - for 
general reviews see \citet{PayneTAAJ92} or \citet{Castep}. For the results 
shown a planewave basis set was used to represent the wavefunction, with a 
$400\,\mbox{eV}$ cutoff and $0.05\,\Ang^{-1}$ sampling of reciprocal space 
using the Monkhorst-Pack scheme \citep{MonkhorstP76,PackM77}. Non-local 
ultrasoft psuedopotentials \citep{Vanderbilt90} were used to describe the 
electron-ion interaction and the local density approximation (LDA) 
\citep{CeperleyA80} was used to treat exchange correlation effects. The LDA 
was used in preference to the various generalized gradient approximation 
(GGA) functionals, as these are known to underbind weakly interacting systems 
such as graphite sheets. The positions of the ions and the unit cell vectors 
were simultaneously optimized using the quasi-Newton method of 
\citet{PfrommerCLC97}, with the reciprocal space sampling density and 
effective planewave cutoff energy maintained at all times.

Only update by roulette wheel selection is used when performing a polymorph 
search to allow a large amount of variation in the cell bond lengths and 
angles, and in the ionic positions. Out of 8 population members which had all 
been randomly initialized, at the end of the tenth generation there were five 
which had a graphite-like character and three which had a Lonsdaleite 
character. A summary of these results, compared with diamond, graphite and 
Lonsdaleite, is shown in table \ref{tab:c4}. Structure 1 is shown in more 
detail in figure \ref{fig:c4_01}.  Within the LDA the binding between 
graphene sheets in graphite is only very weakly dependent on the alignment of 
the sheets, and thus any alignment is permissible, as seen in the differences 
of alignment between figure \ref{fig:c4_01} and figure \ref{fig:c4_05} which 
shows structure 5. Figure \ref{fig:c4_07} shows structure 7 which is 
Lonsdaleite. Since this is a snapshot of the population after 10 generations, 
any configurations of ions are possible, and so diamond structures need not 
be expected. We are currently optimizing the algorithm and the values of the 
parameters in order to generate as wide a range of structures as possible and 
results will be presented in a forthcoming publication \citep{AbrahamP06}.

\section{Conclusions}

We have demonstrated a general method for first principles determination of 
crystal structures using a genetic algorithm in a periodic supercell.
This technique exploits the inherent periodicity in the system in the 
calculation of crossover between parent members of the population by using a 
periodic cut in the crossover operation. This shows much faster convergence 
when compared to crossover performed using a planar cut, as used in lower 
dimensional studies.

\begin{table}[b]
\begin{ruledtabular}
\begin{center}
\begin{tabular}{c c c c}
Structure & Categorization  & \multicolumn{2}{c}{Bond Lengths}\\
\hline
Diamond     & -                & $1.52672\,\Ang$ & $1.52672\,\Ang$ \\
            &                  & $1.52672\,\Ang$ & $1.52672\,\Ang$ 
\xstrut\\
Graphite    & -                & $1.40731\,\Ang$ & $1.40731\,\Ang$
\xstrut\\
Lonsdaleite & -                & $1.52244\,\Ang$ & $1.52244\,\Ang$ \\
            &                  & $1.54583\,\Ang$ & $1.54583\,\Ang$ 
\xstrut\\
1           & Graphite-like    & $1.40696\,\Ang$ & $1.40699\,\Ang$
\xstrut\\
2           & Lonsdaleite-like & $1.52241\,\Ang$ & $1.52241\,\Ang$ \\
            &                  & $1.54583\,\Ang$ & $1.54585\,\Ang$ 
\xstrut\\
3           & Graphite-like    & $1.40679\,\Ang$ & $1.40699\,\Ang$ 
\xstrut\\
4           & Graphite-like    & $1.40707\,\Ang$ & $1.40721\,\Ang$ 
\xstrut\\
5           & Graphite-like    & $1.40701\,\Ang$ & $1.40715\,\Ang$ 
\xstrut\\
6           & Lonsdaleite-like & $1.52225\,\Ang$ & $1.52236\,\Ang$ \\
            &                  & $1.54602\,\Ang$ & $1.54629\,\Ang$ 
\xstrut\\
7           & Lonsdaleite-like & $1.52225\,\Ang$ & $1.52240\,\Ang$ \\
            &                  & $1.54569\,\Ang$ & $1.54599\,\Ang$ 
\xstrut\\
8           & Graphite-like    & $1.40736\,\Ang$ & $1.40746\,\Ang$\\
\end{tabular}
\caption{Summary of the eight carbon structures found in polymorph search 
compared with diamond, graphite and Lonsdaleite.}
\label{tab:c4}
\end{center}
\end{ruledtabular}
\end{table}

There are a number of advantages to this method. First, by  performing all 
crossovers in fractional co-ordinates each population member may be allowed 
to have unit cells which have different sizes and shapes, and if the local 
minimizer also optimizes the cell vectors then the optimization process will 
not be biased by choice of initial structure. Indeed, we always start from 
an initial random structure so there is no initial bias to any preconceived 
solution. Secondly, we also suggest that the roulette selection method could 
be used as a polymorph search technique, since this does not force the 
population down into a single basin, but allows the search space to be 
continually explored. Future work will include further {\it ab initio}
studies using a variable number of atoms and the application of this method 
to surfaces. A systematic study of the ab initio carbon polymorphs with the 
genetic algorithm discussed is still in progress and will be presented 
in a forthcoming publication \citep{AbrahamP06}.

\section{Acknowledgments}

The authors would like to thank Prof. Rex Godby for suggesting the use of a 
periodic cut. Calculations were performed on our departmental Beowulf 
cluster, EPSRC grant R47769 from the Multi-Project Research Equipment 
Initiative. NLA is grateful to the EPSRC for financial support.

\end{document}